\documentclass{article}

\usepackage[english]{babel}
\usepackage{listings}

\usepackage[letterpaper,top=2cm,bottom=2cm,left=3cm,right=3cm,marginparwidth=1.75cm]{geometry}
\usepackage[backend=bibtex, style=numeric, sorting=none]{biblatex}
\usepackage{amsmath}
\usepackage{graphicx}
\usepackage{authblk}
\usepackage{float}
\usepackage{csquotes}
\usepackage[dvipsnames]{xcolor}
\usepackage[colorlinks=true, allcolors=blue]{hyperref}

\title{PlayMolecule Viewer: a toolkit for the visualization of molecules and other data}
\author[+,1]{Mariona Torrens-Fontanals}
\author[+,1]{Panagiotis Tourlas}
\author[1]{Stefan Doerr}
\author[*,2,3,4]{Gianni De Fabritiis}
\affil[1]{Acellera Labs, C Dr Trueta 183, 08005, Barcelona, Spain}
\affil[2]{Computational Science Laboratory, Universitat Pompeu Fabra, Barcelona Biomedical Research Park (PRBB), C Dr.  Aiguader 88, 08003, Barcelona, Spain}
\affil[3]{Acellera, Devonshire House 582 Honeypot Lane Stanmore Middlesex, HA7 1JS United Kingdom}
\affil[4]{Institució Catalana de Recerca i Estudis Avançats (ICREA), Passeig Lluis Companys 23, 08010 Barcelona, Spain}
\affil[+]{Equal contribution}
\affil[*]{E-mail: \href{mailto:g.defabritiis@acellera.com}{g.defabritiis@acellera.com}}

\date{}

\begin{document}
\maketitle

\begin{abstract}
PlayMolecule Viewer is a web-based data visualization toolkit designed to streamline the exploration of data resulting from structural bioinformatics or computer-aided drug design efforts. By harnessing state-of-the-art web technologies such as WebAssembly, PlayMolecule Viewer integrates powerful Python libraries directly within the browser environment, which enhances its capabilities of managing multiple types of molecular data. With its intuitive interface, it allows users to easily upload, visualize, select, and manipulate molecular structures and associated data. The toolkit supports a wide range of common structural file formats and offers a variety of molecular representations to cater to different visualization needs. PlayMolecule Viewer is freely accessible at \href{https://open.playmolecule.org/}{open.playmolecule.org}, ensuring accessibility and availability to the scientific community and beyond.
\end{abstract}

\section{Introduction}
Molecular visualization is an essential tool for computational chemists and biologists, enabling them to explore and understand complex protein structures and unravel vital insights for drug discovery \cite{ Mura2010, Olson2018, Donoghue2010}. The visual inspection of three-dimensional models helps to understand the relationship between protein structure and function. It also allows exploring protein active sites and functional hot spots that can be targeted in drug design, among others.

Traditionally, molecular viewers have predominantly existed as desktop applications due to the demanding nature of 3D graphics \cite{Chavent2011,PyMOL, Humphrey1996, Krieger2014, Pettersen2004}. However, their accessibility and applicability are often hindered by the need to install specialized software. In contrast, web-based tools have emerged as a valuable alternative for the analysis of macromolecular structures  \cite{Rose2015, Sehnal2021, Rego2015, JmolRef, REYNOLDS20182244, Bekker2016, Wang2020,Sehnal2017}. Web-based molecular viewers offer several advantages, such as platform independence and no software installation requirements. These viewers provide a user-friendly and accessible platform, enabling scientists and non-scientific communities alike to engage with molecular visualization and expand the reach of protein research.

The emergence of WebAssembly \cite{WebAssemblyCoreSpecification1}, a web technology that allows running code written in multiple languages in any modern browser, has significantly expanded the capabilities of web-based applications. This includes the ability to run scripts written in interpreted languages such as Python directly within the browser \cite{Pyodide}. WebAssembly holds great potential for advancing web applications in the realm of molecular sciences \cite{Wang2022, Abriata2017, Grzesik2022}, with several tools already using this technology in their development \cite{Jiang2016, Kochnev2020, Kochnev2022, RDKitJs}.

In this context, we introduce PlayMolecule Viewer, a web-based data visualization toolkit designed to facilitate the visualization and comprehension of results from structural bioinformatics or computer-aided drug design efforts. This toolkit allows users to easily upload, visualize, select and manipulate molecular structures and associated data.  By incorporating WebAssembly \cite{WebAssemblyCoreSpecification1}, it leverages the strengths of Python libraries directly within the browser environment. This enhances its capabilities in managing multiple data types, ranging from static and dynamic structures to data tables and plots. This versatility allows users to integrate their data into PlayMolecule Viewer, gaining valuable insights from their molecular structures.

\section{Features}

\subsection{Visualization of molecule structures}
PlayMolecule Viewer provides comprehensive support for loading various types of molecule data, including protein structures and small molecules, molecular dynamics trajectories, and volumetric data. All the main file formats are supported (see section \ref{subsec:supported_formats}). This flexibility enables researchers to analyze a wide range of structural data within the application.

\begin{figure}
\centering
\includegraphics[width=1\linewidth]{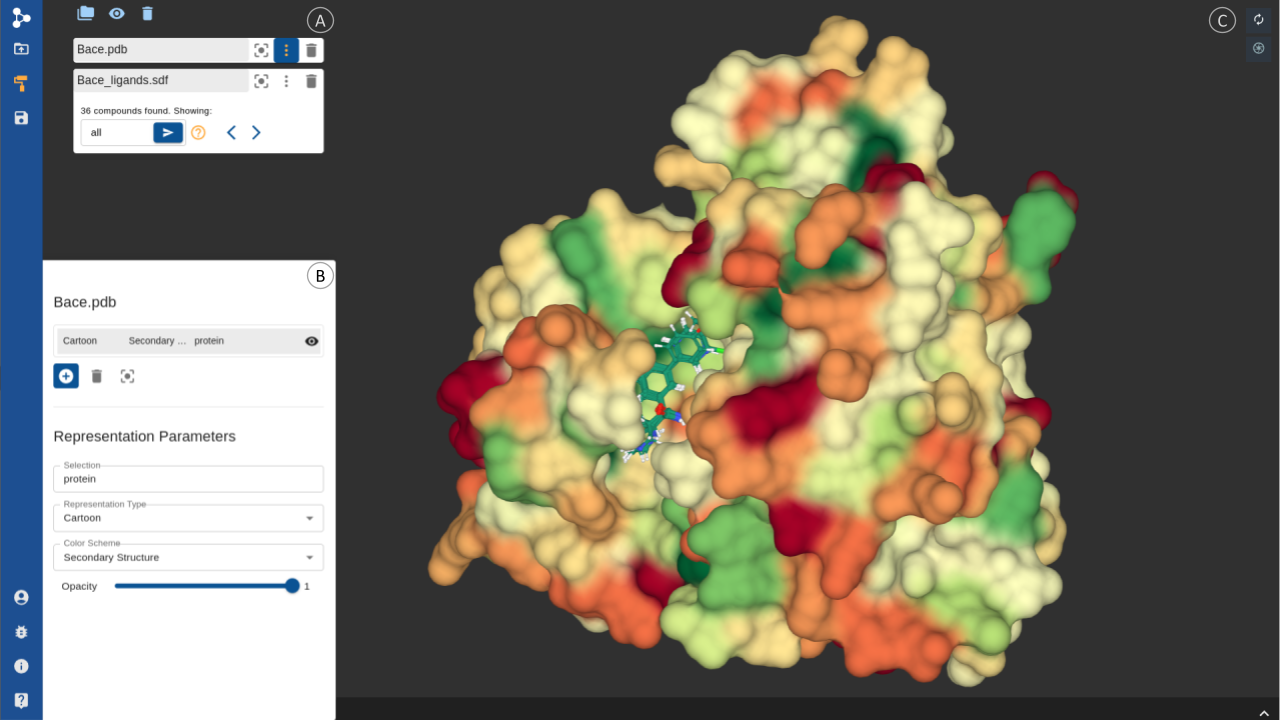}
\caption{Screenshot of  PlayMolecule Viewer showcasing molecular structure controls and customizable representations.
(A) The "3D viewer controls" panel displays the list of loaded molecular structures and provides options to fine-tune their display.
The top bar menu controls grouping, visibility, and deletion.
(B) The representations menu allows the creation and customization of molecular representations. Users can define their selections and choose from various representation types, color schemes, and opacity levels.
(C) The resulting molecular representations are rendered within the viewport using the mol* toolkit \protect{\cite{Sehnal2021}}, providing users with an interactive visualization experience. 
}\label{fig:PMV_structures}
\end{figure}

\begin{figure}
\centering
\includegraphics[width=1\linewidth]{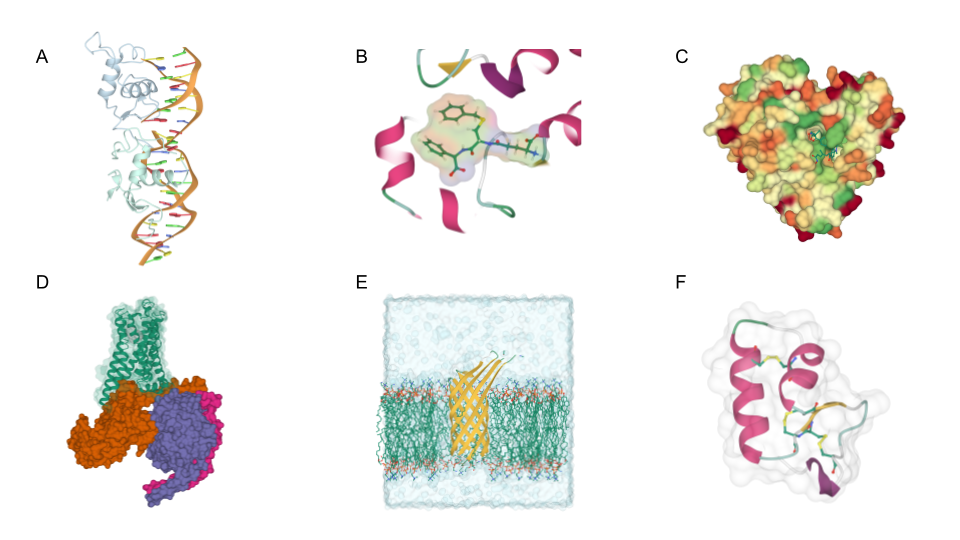}
\caption{Molecular visualizations generated with PlayMolecule Viewer, based on the mol* visualization toolkit \protect{\cite{Sehnal2021}}, illustrating different molecular selections, representation types, coloring schemes and opacity levels. (A) Protein receptor-DNA complex (PDB ID 1a6y), (B) ligand in its binding pocket (PDB ID 10gs), (C) molecular surface of a protein-ligand complex (PDB ID 1t64), (D) protein-protein interaction (PDB ID 8e9y), (E) molecular system including a protein embedded in a cellular membrane and water (PDB ID 3qrc), (F) disulfide bridges (PDB ID 1crn)}\label{fig:representation_examples}
\end{figure}

Loaded structures are displayed in the "3D Viewer Controls" panel (Fig. \ref{fig:PMV_structures}). This panel allows users to efficiently organize their uploaded structures into groups, focus the visualization on specific structures of interest, and control their representations. 
Thanks to the implementation of the mol* toolkit \cite{Sehnal2021}, PlayMolecule Viewer offers a diverse set of molecular structure representations, such as \emph{ball and stick}, \emph{spacefill}, or \emph{cartoon}, that can be combined to create complex and informative molecular views.
Additionally, multiple color schemes are available, providing further visualization options for the structures.
Users can also select and control the representations of structure subsets (see section \ref{subsec:making_selections}), enhancing their ability to focus on and analyze particular regions of interest (Fig. \ref{fig:representation_examples}).

Depending on the file type, additional options are provided. For example, when working with trajectories, users can play and pause the animation, while for volumetric data, they can adjust the isovalue to visualize desired regions. Additionally, when dealing with collections of small molecules, users have control over the visualization of each compound.

\subsection{Making selections}\label{subsec:making_selections}
Making selections is a fundamental aspect of data visualization, enabling users to create compelling, informative, and complex visualizations. For that, PlayMolecule Viewer uses the VMD selection language \cite{Humphrey1996}, which is based on keywords with associated values (Table \ref{table:atom_sel}).

\begin{table}
\caption{Common atom selection keywords. Please note that this table provides a representative selection of keywords, and many more are supported by the platform.}
\label{table:atom_sel}
\begin{tabular}{l l l} 
    \hline
  Keyword & Type & Description \\ 
  \hline
all &	bool 	& Everything  \\ 
name & 	str & 	Atom name \\
type & 	str & 	Atom type \\
index & 	num & 	Atom number, starting at 0 \\
serial & 	num & 	Atom number, starting at 1 \\
element & 	str & 	Atomic element symbol \\
chain & 	str & 	One-character chain identifier \\
protein & 	bool & 	Protien atoms \\
nucleic & 	bool & 	Nucleic acid atoms \\
backbone & 	bool & 	Backbone of a protein or nucleic acid. \\
sidechain & 	bool & 	Non-backbone atoms and bonds \\
water & 	bool & 	All water atoms \\
resname & 	str & 	Residue name \\
resid & 	num & 	Residue id \\
segname & 	str & 	Segment name \\
x, y, z 	& float 	& x, y, or z coordinates \\ [0.5ex] 
 
\end{tabular}
\end{table}

Atoms can be selected on the basis of a boolean keyword, such as \textit{protein}, \textit{water}, and \textit{backbone}. For instance, the selection \textit{"protein"} will display all protein atoms. Selections can also be formulated as a keyword followed by a list or a range of values. Some examples are \textit{"name CA"} (all atoms with the name CA), \textit{"resname ALA PHE ASP"} (all atoms in either alanine, phenylalanine, or asparagine) or \textit{"resid 50 to 100"} (atoms with residue index from 50 to 100).

Selections can also be combined with the boolean operators \textit{and} and \textit{or}, and modified by \textit{not}. For instance,  \textit{“protein and chain A and not name H”} corresponds to all protein atoms in chain A that are not hydrogen.

Finally, advanced criteria such as regular expressions (e.g. \textit{"name 'C.*' "} to select all atoms with names starting with \textit{C}) and spacial selection (e.g. \textit{"same residue as within 5 of resname BEN"} to select residues within 5 \r{A} of the residues named BEN) are also accepted, adding an extra layer of versatility to the selection processes. 

More information on atom selection can be found at \href{https://open.playmolecule.org/docs/selection}{open.playmolecule.org/docs/selection}.

\subsection{Integration of Tabular Data}
\begin{figure}
\centering
\includegraphics[width=1\linewidth]{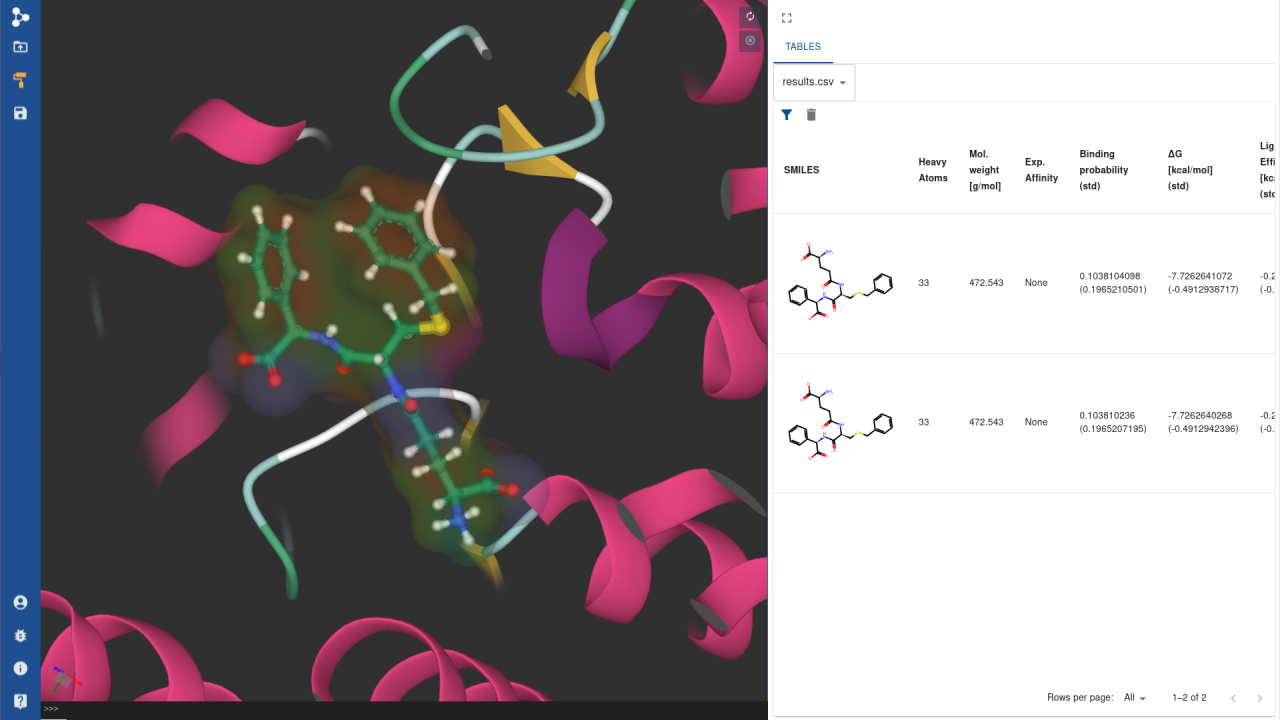}
\caption{Screenshot of PlayMolecule Viewer displaying the "Data Visualization" panel, which provides a powerful tool for exploring tabular data.
The panel's top menu allows users to navigate through all loaded tables and filter the displayed columns. Table rows can be sorted based on a specific column by simply clicking on the column header. Tables prove particularly valuable when working with small molecules. Notably, if the table includes a column labeled "SMILES," the viewer automatically converts the SMILES representations into 2D depictions.
}\label{fig:PMV_table}
\end{figure}

In addition to molecular structures, PlayMolecule Viewer allows users to upload tables in CSV format (Fig. \ref{fig:PMV_table}). This feature is useful, among others, when working with libraries of small molecules. Notably, if the table includes a column labeled "SMILES", the application automatically converts the SMILES notation into a visual depiction of the small molecule.

Moreover, the tables generated offer interactive features. For instance, users can enrich the table by adding a column named "Path" containing the filenames of loaded molecular structures. When a table row is clicked, the viewer automatically focuses on the corresponding structures, facilitating the navigation between tabular data and their associated molecular representations. This interactivity enhances the user experience and streamlines the understanding and exploration of complex molecular datasets.

\subsection{Integration of Static and Interactive Plots}

\begin{figure}
\centering
\includegraphics[width=1\linewidth]{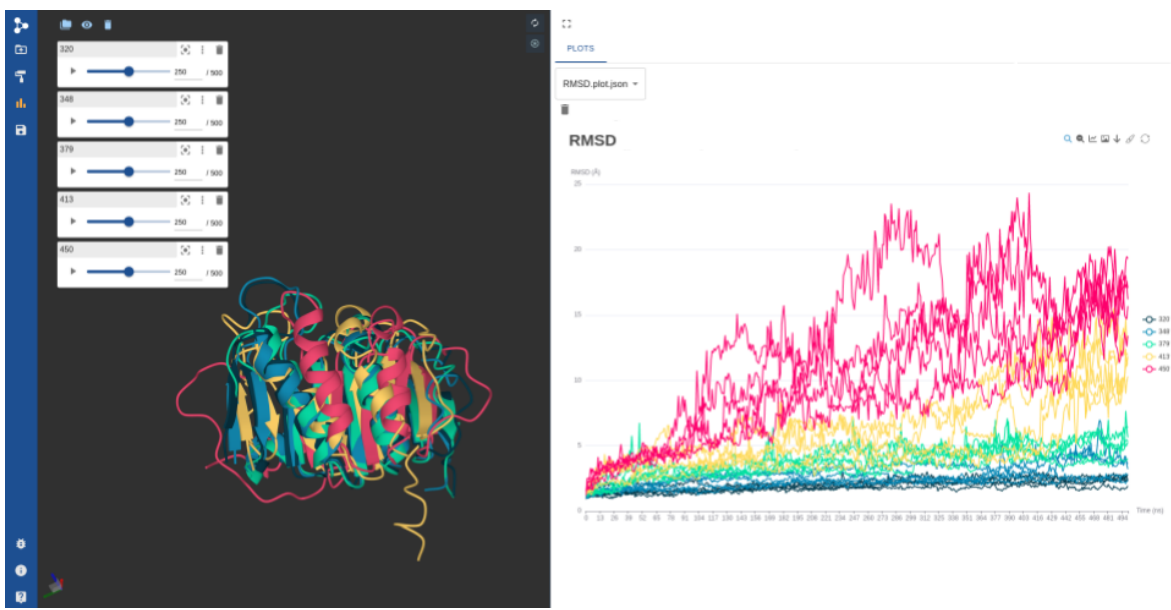}
\caption{ PlayMolecule Viewer offers support for interactive plots, helping users to gain deeper insights into their structural data. For instance, users can visualize time-dependent parameters, such as Root Mean Square Deviation (RMSD), alongside their associated molecular dynamics simulations.
}\label{fig:PMV_plots}
\end{figure}

PlayMolecule Viewer offers the visualization of both static or interactive plots, allowing users to integrate complex data representations with their molecular structures  (Fig. \ref{fig:PMV_plots}). 
Users can upload static plots as plain images. Moreover, they can display interactive plots by uploading a configuration file in the format of Apache ECharts Echarts \cite{LI2018136}, a popular JavaScript charting library that provides a wide range of chart types and customization options, covering from line and scatter plots to heat maps and graphs. For its correct recognition, the plot configuration file should have the extension .plot.json.

These data visualization capabilities can be useful to display time-dependent properties of molecular systems alongside a molecular dynamics trajectory, providing a comprehensive view of dynamic behaviors. Additionally, displaying energy landscapes or binding affinity profiles together with protein structures can aid in the understanding of protein conformations.

\subsection{Python Terminal for Interactive Molecular Analysis}
Thanks to the implementation of Pyodide \cite{Pyodide}, a Python distribution for the browser based on WebAssembly \cite{WebAssemblyCoreSpecification1}, PlayMolecule Viewer includes a Python terminal that allows users to interact with the loaded molecules using Moleculekit \cite{Doerr2016}. Users can execute Python code within the terminal, enabling them to obtain information about the molecules, apply filters, perform calculations, and more. This powerful feature enhances the flexibility and interactivity of the viewer, allowing users to gain deeper insights into their molecules without leaving the application.

For instance, we can access a specific system loaded in PlayMolecule Viewer by employing the \textbf{get\_system(system\_index)} function, where \textbf{system\_index} corresponds to the 0-based position of the system within the viewer. This returns a Moleculekit molecule object, which grants us access to the extensive functionalities of this library. Thus, we can retrieve specific properties of the system, such as the sequence or the protein residues in contact with the ligand. In the case of trajectories, users can extract additional data, such as the number of frames:

\begin{lstlisting}
mol = get_system(0)
mol.sequence()
mol.get('resid', sel='name CA and same residue as within 4 of resname MOL')
mol.numFrames
\end{lstlisting}

Moreover, it is possible to modify the system. For example, users can apply filters or translations and align to another molecule:

\begin{lstlisting}
# Remove water molecules:
mol.filter("not water")
# Translate to the origin or coordinates:
import numpy as np
coo = mol.get('coords')
c = np.mean(coo, axis=0)
mol.moveBy(-c)
# Structural alignment
mol.align("name CA", refmol=mol2)
\end{lstlisting}

\subsection{Save and Load Viewer Sessions}\label{subsec:session_save}
To facilitate work continuity, PlayMolecule Viewer enables users to save their session state by clicking on "Save Viewer State." This feature captures in a session file the current configurations, including the chosen representations, selections, and any modifications made to the structures. Users can later reload the session file, restoring their previous visualization settings and ensuring a seamless continuation of their analysis.

\subsection{Data formats accepted by PlayMolecule Viewer} \label{subsec:supported_formats}
PlayMolecule Viewer is currently able to load and visualize multiple file formats:
\begin{itemize}
  \item Topology and coordinate files: prmtop, prm, psf, mae, mol2, gjf, xyz, pdb, ent, pdbqt, GROMACS \cite{Berendsen1995} top, crd, cif, rtf, prepi, sdf, mmtf, mmtf.gz, and coor.
  \item Volumetric data, such as molecular densities: cube
  \item Trajectories: xtc, xsc, netcdf, dcd, trr, and binpos  
  \item Tables: csv
  \item Interactive plots: Echarts \cite{LI2018136} configuration path with extension plot.json.
  \item Static plots or other images: png and svg
\end{itemize}

\section{Implementation}
PlayMolecule Viewer is developed using TypeScript \cite{TypeScript} and the React library \cite{React}, which provide a solid foundation for building a robust and user-friendly web application. 

Leveraging the power of Pyodide \cite{Pyodide}, this toolkit can execute Python code directly within the browser environment, which enables the integration of powerful Python libraries for efficient data manipulation.
This is the case of Moleculekit \cite{Doerr2016}, a Python library for the manipulation of biomolecular structures.
Using Moleculekit \cite{Doerr2016}, PlayMolecule Viewer can accept a wide range of file formats which ensures compatibility with various molecular data sources and enhances the application's flexibility.
Additionally, Pandas \cite{reback2020pandas} is employed to generate tables based on CSV files.
The creation of molecular representations based on the loaded structures is handled by the mol* \cite{Sehnal2021} visualization toolkit. Moreover, we use RDKitJS \cite{RDKitJs} to generate depictions of small molecules based on SMILES.

Due to the resource-intensive nature of these tools and the single-thread design of the JavaScript language,
PlayMolecule Viewer makes use of the WebWorker \cite{WebWorkerSpec} API, which allows long-running background tasks, such as the Python interpreter, to execute independently from the user interface thread.
Overall, this modular architecture provides PlayMolecule Viewer with a user-friendly interface and wide data processing capabilities. 
 
\section{PlayMolecule Viewer as a standalone application}
In addition to the web platform, PlayMolecule Viewer is also available as a standalone application. This version brings several notable advantages compared to its web-based counterpart, including offline accessibility and enhanced performance. The standalone application benefits from direct access to the user's local file system, simplifying the process of loading files into the viewer. Users have the option to load files through command-line operations, reducing the steps necessary to input data. For example, \emph{pmview "directory/*.pdb"} will open all PDB files in a given directory. 

The PlayMolecule Viewer standalone application can be installed using conda (\texttt{conda install pmview -c acellera}).

\section{Comparative overview: PlayMolecule Viewer and other browser-based molecular viewers} 
Numerous web-based molecular viewers have been developed to date \cite{Rose2015, Sehnal2021, Rego2015, JmolRef, REYNOLDS20182244, Bekker2016, Wang2020,Sehnal2017}. Most function primarily as embeddable libraries, while also offering full web application versions. Focusing on web applications, a majority of these tools primarily support static structures, but not molecular dynamics trajectories. To our knowledge, besides PlayMolecule Viewer, to this date only mol* Viewer \cite{Sehnal2021} and NGL Viewer \cite{Rose2015} (on which mol* is built) support the visualization of molecular dynamics trajectories uploaded by the user.  Mol* Viewer and NGL Viewer offer an extensive range of options for visualizing and inspecting molecular structures, surpassing PlayMolecule Viewer in certain aspects. For instance, mol* Viewer provides features like sequence view, a wider array of customization options, and advanced rendering capabilities. In contrast, PlayMolecule Viewer prioritizes ease of use, featuring a simplified interface and comprehensive, accessible documentation.

A significant distinction of PlayMolecule Viewer from other existing online viewers lies in its WebAssembly \cite{WebAssemblyCoreSpecification1} implementation, enabling Python execution within the browser. This implementation significantly broadens its capability to handle different types of molecular data. Particularly, it facilitates the support of multiple structure file formats, thanks to the Moleculekit library \cite{Doerr2016}, as well as enabling the integration of interactive tables and plots, thanks to libraries such as Pandas \cite{reback2020pandas} and Numpy \cite{harris2020array}. While some viewers support the integration of annotations such as protein domains and single nucleotide variations \cite{Wang2020, Bekker2016}, few, if any, support the visualization of user-generated tables and plots, thereby providing flexibility for integration with other analysis results. Additionally, PlayMolecule Viewer includes a Python console, a unique feature for interacting with and analyzing the uploaded systems. While some online viewers feature consoles for modifying molecular visualizations \cite{Wang2020, JmolRef, Bekker2016}, as far as our knowledge extends, none include a dedicated Python console.

Overall, the strengths of PlayMolecule Viewer lie in its beginner-friendly interface, WebAssembly implementation allowing Python execution in the browser, and the incorporation of a Python console for system interaction and analysis, setting it apart from other browser-based molecular viewers.

\section{Conclusions}
By leveraging cutting-edge web technologies such as WebAssembly \cite{WebAssemblyCoreSpecification1}, PlayMolecule Viewer emerges as a powerful and versatile web application for interactive molecular data visualization. Accessible through a user-friendly interface, it allows users to explore a wide range of structure file formats, including molecular dynamics simulations, as well as data tables and plots. With a rich set of molecular representation types and color schemes, together with selection and superimposition capabilities, users can create complex molecular views. The ability to save visualization states ensures easy recovery of previous work. Notably, the toolkit incorporates a Python terminal, enabling users to interact with loaded molecules using the Moleculekit \cite{Doerr2016} Python library. PlayMolecule Viewer is freely available at \href{https://open.playmolecule.org/}{open.playmolecule.org}.

\section{Data and Software Availability}
PlayMolecule Viewer is freely accessible to all users at \href{https://open.playmolecule.org/}{open.playmolecule.org}. The terms of use can be found at \href{https://open.playmolecule.org/docs/termsofuse}{open.playmolecule.org/docs/termsofuse}.

The standalone application can be installed via \texttt{conda install pmview -c acellera}. Creating an exclusive conda environment for the viewer prior to the installation is recommended.

\section{Acknowledgments}
The project PID2020-116564GB-I00 has been funded by MCIN/ AEI /10.13039/50110001103. Grant PTQ2021-011670 funded by MCIN/AEI/ 10.13039/501100011033. This project has received funding from the European Union’s Horizon 2020 research and innovation programme under grant agreement No 823712.

\printbibliography

\end{document}